\documentclass[a4paper]{article}

\usepackage{19lomcon}        
\usepackage{cite}             
\usepackage{epsfig}           
\usepackage{epstopdf}

\begin{document}


\title{SUBGROUP OF FINITE RENORMALIZATIONS, CONSERVING THE NSVZ RELATION IN $\mathcal{N} = 1$ SQED}

\author{I.O.Goriachuk \email{io.gorjachuk@physics.msu.ru} and
        A.L.Kataev \email{kataev@ms2.inr.ac.ru}
        }

\affiliation{$^{a}$ Faculty of Physics, Moscow State University, 119991
Moscow, Russia}

\affiliation{$^{b}$ Institute for Nuclear Research
of the Russian Academy of Sciences, 117312, Moscow, Russia}


\date{27.08.19}
\maketitle


\begin{abstract}
In supersymmetric quantum electrodynamics there is the exact relation between the expressed through 
the unrenormalized coupling 
gauge beta-function and the anomalous dimension of matter superfields.
In the present report we describe the subgroup of general renormalization group transformations,
conserving this relation exactly in all orders of the perturbation theory
in terms of the renormalized coupling constant.
\end{abstract}

\vspace*{-6.0cm}

\begin{flushright}
INR-TH-2019-026
\end{flushright}

\vspace*{5.0cm}

\section{Introduction}
The calculations of quantum corrections in various field theory models are  leading to 
the appearance of  the  ultraviolet
divergences, which can be removed from physical quantities by the renormalization procedure
\cite{Vladimirov:1979ib}.
Renormalization transformations constitute the renormalization group \cite{Bogolyubov:1980nc}.
In the present report we construct the subgroup of these transformations, conserving the validity
of  the exact Novikov, Shifman, Vainshtein and Zakharov (NSVZ)  relation 
\cite{Novikov:1983uc,Jones:1983ip} for the renormalized $\beta$-function and anomalous dimension of the 
matter superfields in
$\mathcal{N} = 1$ supersymmetric  (SUSY) quantum electrodynamics (SQED) with $N_f$ flavours.
In this SUSY model  this relation  \cite{Vainshtein:1986ja} is :
\begin{equation}
\label{eq:NSVZ}
\frac{\widetilde{\beta}\left(\alpha\right)}{\alpha^2} =
\frac{N_f}{\pi}\left(1 - \widetilde{\gamma}\left(\alpha\right)\right).
\end{equation}
where $\alpha=e^2/(4\pi)$. 
This formula is written for the renormalization group functions, defined in terms
of the renormalized coupling constant in the following standard way:
\begin{equation}
\label{eqs:RGfs}
\widetilde{\beta}\left(\alpha\right) =
\left. \frac{d\alpha\left(\alpha_0, \Lambda/\mu\right)}{d\ln{\mu}}\right\arrowvert_{\alpha_0}, \quad
\widetilde{\gamma}\left(\alpha\right) =
\left. \frac{d\ln{Z\left(\alpha\left(\alpha_0, \Lambda/\mu\right), \Lambda/\mu\right)}}{d\ln{\mu}}
\right\arrowvert_{\alpha_0},
\end{equation}
where $\Lambda$ is the dimensionful regularization parameter, $\alpha$ is the renormalized coupling constant and 
$Z$ is the renormalization constant of the chiral   matter superfields $\phi_i$ (encoding the quantum renormalization of these superfields). Note, that both 
$\alpha$ and $Z$ are defined at a particular normalization scale
$\mu$:
\begin{equation}
\alpha = \alpha\left(\alpha_0, \Lambda/\mu\right), \quad
\phi_{i} = Z^{\frac{1}{2}}\left(\alpha, \Lambda/\mu\right) \phi_{Ri}, \quad
\phi^{*}_{i} = Z^{\frac{1}{2}}\left(\alpha, \Lambda/\mu\right) \phi_{Ri}^{*}.
\end{equation}
The total derivatives with respect to the logarithm of this scale act at a fixed value of the bare
coupling constant $\alpha_0$.
These beta- and gamma-functions depend on the choice of the renormalization prescription
\cite{Vladimirov:1979my,Kataev:2014gxa}.
Therefore, the NSVZ formula   is valid only with certain renormalization prescriptions
(called the NSVZ schemes), which are related by specific finite renormalization (FR) procedures.

\section{Specific NSVZ schemes}
The first  scheme in which the NSVZ relation (\ref{eq:NSVZ}) for the 
renormalization group functions defined in terms of the renormalized coupling constant is valid 
was  constructed  in all orders of the perturbation theory,
when the higher derivatives (HD) regularization \cite{Slavnov:1971aw,Slavnov:1972sq}
was applied and  the renormalization procedure was based  on
minimal subtraction of $\ln(\Lambda/\mu)$-terms (MSL)\cite{Kataev:2013eta}. Within the MSL-scheme 
the renormalized NSVZ relation  seems to be true for the non-Abelian theories 
in all orders of perturbation theory as well \cite{Stepanyantz:2016gtk}.
This HD-based scheme is constructed  to insure the coincidence of the N=1 SUSY QED renormalization 
group functions defined in terms of renormalized and bare couplings \cite{Kataev:2013eta} and on the 
fact that within HD-regularization the bare coupling constant-dependent  NSVZ equation 
is satisfied in all orders of perturbation theory.  In the Abelian case it was demonstrated in \cite{Stepanyantz:2011jy,Stepanyantz:2014ima}, while for non-Abelian theories this was confirmed by numerous multiloop calculations (see, e.g., \cite{Kazantsev:2018nbl}).
Besides, the HD regularization is mathematically consistent and has a number of attractive features.
It can be formulated in a manifestly supersymmetric way \cite{Krivoshchekov:1978xg, West:1985jx}
and provides the factorization of loop integrals for the expressed through 
unrenormalized couplings  beta-functions of N=1 SUSY theories 
into the integrals from 
double total derivatives \cite{Stepanyantz:2019ihw}. This important feature simplifies essentially the
concrete multi-loop calculations in the entering into NSVZ relations quantities both in the 
Abelian and non-Abelian SUSY cases.

However, it is known, that the NSVZ relation does not take place, when the theory is regularized with the help of
dimensional reduction (DRED) and renormalized according to the $\overline{\mbox{DR}}$ subtraction prescription
\cite{Jack:1996vg,Jack:1996cn}.
In this case the equation (\ref{eq:NSVZ}) is not valid already in the lowest scheme-dependent 3-loop order
for the $\beta$-function.
Nevertheless, it is always possible to make a FR of $\alpha$, restoring the NSVZ relation.
This FR, required for the $\overline{\mbox{DR}} + \mbox{NSVZ}$ scheme, can be reformulated in terms of certain
boundary conditions imposed on the renormalization constants \cite{Aleshin:2016rrr},
at least, at the 3-loop level.

The third important ultraviolet renormalization presription, which was used in SQED,  is the 
scheme with the subtractions  on the mass shell, where the renormalized mass coincides 
with the "physical" one, defined as pole in the renormalized propagators of the matter 
superfields \cite{Smilga:2004zr}.
Recently it was shown that the application of the  on-shell scheme  in SQED (regularized by HD)  gives  the exact NSVZ
$\beta$-function  in all orders . Therefore, the on-shell prescription is distinguished 
not only from phenomenological reasons, as in non-supersymmetric  QED  \cite{Broadhurst:1992za},  but from 
theoretical reasons as well \cite{Kataev:2019olb}.  

It is known, that  all these three different NSVZ renormalization prescriptions provide different answers
for the SQED  $\beta$-function starting from the 3-loop level. 
All of these NSVZ schemes belong to a particular class
 of the related by FR  renormalization prescriptions, which respect the SQED NSVZ relations
\cite{Goriachuk:2018cac}.
The explicit expressions for these FRs in the lowest orders of the perturbation theory can be found in
 \cite{Kataev:2019olb} and \cite{Goriachuk:2018cac}.

\section{The class of NSVZ schemes}
Finite renormalization in its general form is the following redefinition of the renormalized
coupling constant $\alpha$ and the $Z$-factor for the renormalization of matter:
\begin{equation}
\label{eqs:fin_ren}
\alpha^\prime = \alpha^\prime\left(\alpha\right), \quad
Z^\prime\left(\alpha^\prime,~\Lambda/\mu\right) =
z\left(\alpha\right)Z\left(\alpha,~\Lambda/\mu\right),
\end{equation}
where $\alpha^\prime\left(\alpha\right)$ and $z\left(\alpha\right)$ are arbitrary finite functions.
Under this FR the gauge $\beta$-function and the anomalous dimension $\gamma$ are  changing  as follows:
\begin{equation}
\label{eq:fin_ren_RG}
\widetilde\beta^\prime\left(\alpha^\prime\left(\alpha\right)\right) =
\frac{d\alpha^\prime\left(\alpha\right)}{d\alpha} \widetilde\beta\left(\alpha\right), \quad
\widetilde\gamma^\prime\left(\alpha^\prime\left(\alpha\right)\right) =
\frac{d\ln z\left(\alpha\right)}{d\alpha} \widetilde\beta\left(\alpha\right) +
\widetilde\gamma\left(\alpha\right).
\end{equation}
Let us find the general form of FRs, conserving the NSVZ relation (\ref{eq:NSVZ}).

Consider two arbitrary renormalization schemes, in which the NSVZ relation is valid.
One can always relate them with each other by using  Eq.(\ref{eqs:fin_ren}).
It is possible to show, that this FR should satisfy the condition \cite{Goriachuk:2018cac}:
\begin{equation}
\label{eq:NSVZ_cons_fin_ren}
\frac{1}{\alpha^\prime\left(\alpha\right)} - \frac{1}{\alpha}
- \frac{N_f}{\pi}\ln{z\left(\alpha\right)} = B = - \frac{N_f}{\pi} \ln{\frac{\mu^\prime}{\mu}}.
\end{equation}
where $B$ does not depend on $\alpha$  and reflects the arbitrariness of the choice of the  normalization scale $\mu$
\cite{Vladimirov:1979my}.
This condition is relating the forms of variations of finite functions 
$\alpha^\prime\left(\alpha\right)$,  $z\left(\alpha\right)$ and $B$
and is describing the class of the NSVZ schemes \cite{Goriachuk:2018cac}.
Any FR transformation, converting any  NSVZ scheme into another one, which is belonging to  this class,
should satisfy this condition.
The inverse is also valid: any FR, satisfying (\ref{eq:NSVZ_cons_fin_ren}),
leaves the NSVZ relation valid.
Therefore, only one of the $\alpha$-depended   finite functions
and the parameter  B can be chosen arbitrarily, while  the other function  is  determined unambiguously.

\section{The group and subgroup}
In general finite renormalizations (\ref{eqs:fin_ren}) belong to  the  renormalization group transformations
\cite{Vladimirov:1979ib}, which forms  the  continuously parametric  group \cite{Bogolyubov:1980nc}.
Moreover, they are forming the subgroup of these renormalization group transformations. 
Indeed,  consider two arbitrary FRs with parameters $\{\alpha_1, z_1\}$ and $\{\alpha_2, z_2\}$.
Their  group multiplication, i.e. their  composition 
\begin{equation}
\label{eqs:RG_composit}
\alpha^\prime\left(\alpha\right) = \alpha_2\left(\alpha_1\left(\alpha\right)\right), \quad
z\left(\alpha\right) = z_2\left(\alpha_1\left(\alpha\right)\right) z_1\left(\alpha\right),
\end{equation}
is also a finite renormalization.
One can construct such compositions at an arbitrary order.
This possibility provides associativity of the transformations.
The unit element of this group is the following identical transformation:
\begin{equation}
\label{eqs:RG_unit}
\alpha^\prime\left(\alpha\right) = \alpha, \quad z\left(\alpha\right) = 1.
\end{equation}
For each transformation (\ref{eqs:fin_ren}) there is an inverse one
\begin{equation}
\label{eqs:RG_inverse}
\alpha_{inv}\left(\alpha^\prime\right) = {\alpha^\prime}^{-1}\left(\alpha^\prime\right), \quad
z_{inv}\left(\alpha^\prime\right) =
\frac{1}{z\left({\alpha^\prime}^{-1}\left(\alpha^\prime\right)\right)},
\end{equation}
since the inverse function ${\alpha^\prime}^{-1}$ always exists in perturbative calculations.
It worth to note, that this inverse transformation can be applied both before the initial one
and after it: the resulting composition gives the identical element (\ref{eqs:RG_unit}).
Thus, FR  subgroup satisfy all   properties of a smoothly parameterised
group. They constitute the group of renormalizations in its general sense.
Moreover, the conserving the NSVZ relation FRs constitute the theorerically distinguished in SQED  subgroup, 
which  satisfies 
the condition (\ref{eq:NSVZ_cons_fin_ren}). Let us show this explicitly.

Consider two sequential FRs within the class of NSVZ schemes with parameters $\{\alpha_1, z_1, B^{(1)}\}$
and $\{\alpha_2, z_2, B^{(2)}\}$ respectively, which satisfy the condition (\ref{eq:NSVZ_cons_fin_ren}).
This group multiplication doesn't lead out of the considered subset of transformations.
For each element (\ref{eqs:fin_ren}) of the subset under investigation there is an inverse one, given by
(\ref{eqs:RG_inverse}).
It also belongs to this subset, because it satisfies the condition
\begin{equation}
\frac{1}{\alpha_{inv}\left(\alpha^\prime\right)} - \frac{1}{\alpha^{\prime}} -
\frac{N_f}{\pi} \ln{z_{inv}\left(\alpha^\prime\right)} = B_{inv} 
\end{equation}
with $B_{inv} = - B$. This can be demonstrated by substituting $\alpha^\prime(\alpha)$ into (\ref{eqs:RG_inverse}).
The condition (\ref{eq:NSVZ_cons_fin_ren}) is valid for the identical transformation (\ref{eqs:RG_unit})
with $B = 0$.
Thus, the subset of FRs, distinguished by the condition (\ref{eq:NSVZ_cons_fin_ren}), is the subgroup.

\section{Conclusion}
In this report the subgroup of general renormalization group transformations
is described for $\mathcal{N} = 1$ SQED.
This transformations are finite renormalizations, conserving the NSVZ relation
in terms of the renormalized coupling constant $\alpha$.
They act within the class of the NSVZ renormalization schemes and satisfy the condition
(\ref{eq:NSVZ_cons_fin_ren}).
The considered subgroup is parameterized by a single function $z\left(\alpha\right)$ and a constant $B$.
These results can be generalized to the case of non-Abelian ${\cal N}=1$ supersymmetric gauge theories
with chiral matter superfields.

\section*{Acknowledgments}
We  are grateful to A.E. Kazantsev and K.V. Stepanyantz for valuable discussions,
This work was supported by the Foundation for the Advancement
of Theoretical Physics and Mathematics ''BASIS'', grant No. 17-11-120.



\begin{thebibliography}{99}

\bibitem{Vladimirov:1979ib}
  A.~A.~Vladimirov and D.~V.~Shirkov,
  Sov.\ Phys.\ Usp.\  {\bf 22} (1979) 860
   [Usp.\ Fiz.\ Nauk {\bf 129} (1979) 407].

\bibitem{Bogolyubov:1980nc}
  N.~N.~Bogolyubov and D.~V.~Shirkov,
  {\it ``Introduction To The Theory Of Quantized Fields''}
  (Intersci.\ Monogr.\ Phys.\ Astron.\  {\bf 3}, 1, 1959).

\bibitem{Novikov:1983uc}
  V.~A.~Novikov, M.~A.~Shifman, A.~I.~Vainshtein and V.~I.~Zakharov,
  Nucl.\ Phys.\ B {\bf 229} (1983) 381.

\bibitem{Jones:1983ip}
  D.~R.~T.~Jones,
  Phys.\ Lett.\  {\bf 123B} (1983) 45.

\bibitem{Vainshtein:1986ja}
  A.~I.~Vainshtein, V.~I.~Zakharov and M.~A.~Shifman,
  JETP Lett.\  {\bf 42} (1985) 224
   [Pisma Zh.\ Eksp.\ Teor.\ Fiz.\  {\bf 42} (1985) 182].

\bibitem{Vladimirov:1979my}
  A.~A.~Vladimirov,
  Sov.\ J.\ Nucl.\ Phys.\  {\bf 31} (1980) 558
   [Yad.\ Fiz.\  {\bf 31} (1980) 1083].

\bibitem{Kataev:2014gxa}
  A.~L.~Kataev and K.~V.~Stepanyantz,
  Theor.\ Math.\ Phys.\  {\bf 181} (2014) 1531 [Teor.\ Mat.\ Fiz.\  {\bf 181} (2014) 475].

\bibitem{Slavnov:1971aw}
  A.~A.~Slavnov,
  Nucl.\ Phys.\ B {\bf 31} (1971) 301.

\bibitem{Slavnov:1972sq}
  A.~A.~Slavnov,
  Theor.Math.Phys. {\bf 13} (1972) 1064
   [Teor.\ Mat.\ Fiz.\  {\bf 13} (1972) 174].

\bibitem{Kataev:2013eta}
  A.~L.~Kataev and K.~V.~Stepanyantz,
  Nucl.\ Phys.\ B {\bf 875} (2013) 459.

\bibitem{Stepanyantz:2016gtk}
  K.~V.~Stepanyantz,
  Nucl.\ Phys.\ B {\bf 909} (2016) 316.

\bibitem{Stepanyantz:2011jy}
  K.~V.~Stepanyantz,
  Nucl.\ Phys.\ B {\bf 852} (2011) 71.

\bibitem{Stepanyantz:2014ima}
  K.~V.~Stepanyantz,
  JHEP {\bf 1408} (2014) 096.

\bibitem{Kazantsev:2018nbl}
  A.~E.~Kazantsev, V.~Y.~Shakhmanov and K.~V.~Stepanyantz,
  JHEP {\bf 1804} (2018) 130.

\bibitem{Krivoshchekov:1978xg}
  V.~K.~Krivoshchekov,
  Theor.\ Math.\ Phys.\ {\bf 36} (1978) 745
 [Teor.\ Mat.\ Fiz.\  {\bf 36} (1978) 291].

\bibitem{West:1985jx}
  P.~C.~West,
  Nucl.\ Phys.\ B {\bf 268} (1986) 113.

\bibitem{Stepanyantz:2019ihw}
  K.~V.~Stepanyantz,
  JHEP {\bf 1910} (2019) 011

\bibitem{Jack:1996vg}
  I.~Jack, D.~R.~T.~Jones and C.~G.~North,
  Phys.\ Lett.\ B {\bf 386} (1996) 138.

\bibitem{Jack:1996cn}
  I.~Jack, D.~R.~T.~Jones and C.~G.~North,
  Nucl.\ Phys.\ B {\bf 486} (1997) 479.



\bibitem{Aleshin:2016rrr}
  S.~S.~Aleshin, I.~O.~Goriachuk, A.~L.~Kataev and K.~V.~Stepanyantz,
  Phys.\ Lett.\ B {\bf 764} (2017) 222.

\bibitem{Smilga:2004zr}
  A.~V.~Smilga and A.~Vainshtein,
  Nucl.\ Phys.\ B {\bf 704} (2005) 445

\bibitem{Kataev:2019olb}
  A.~L.~Kataev, A.~E.~Kazantsev and K.~V.~Stepanyantz,
  Eur.\ Phys.\ J.\ C {\bf 79} No. 6 (2019) 477
  
\bibitem{Broadhurst:1992za}
  D.~J.~Broadhurst, A.~L.~Kataev and O.~V.~Tarasov,
  Phys.\ Lett.\ B {\bf 298} (1993) 445

\bibitem{Goriachuk:2018cac}
  I.~O.~Goriachuk, A.~L.~Kataev and K.~V.~Stepanyantz,
  Phys.\ Lett.\ B {\bf 785} (2018) 561.





\end{thebibliography}
\end{document}